# "Life never matters in the DEMOCRATS MIND": Examining Strategies of Retweeted Social Bots During a Mass Shooting Event

**Vanessa L. Kitzie**
*University of South Carolina, USA. kitzie@mailbox.sc.edu*

**Amir Karami**
*University of South Carolina, USA. karami@sc.edu*

**Ehsan Mohammadi**
*University of South Carolina, USA. ehsan2@sc.edu*

**ABSTRACT**

This exploratory study examines the strategies of social bots on Twitter that were retweeted following a mass shooting event. Using a case study method to frame our work, we collected over seven million tweets during a one-month period following a mass shooting in Parkland, Florida. From this dataset, we selected retweets of content generated by over 400 social bot accounts to determine what strategies these bots were using and the effectiveness of these strategies as indicated by the number of retweets. We employed qualitative and quantitative methods to capture both macro- and micro-level perspectives. Our findings suggest that bots engage in more diverse strategies than solely waging disinformation campaigns, including baiting and sharing information. Further, we found that while bots amplify conversation about mass shootings, humans were primarily responsible for disseminating bot-generated content. These findings add depth to the current understanding of bot strategies and their effectiveness. Understanding these strategies can inform efforts to combat dubious information as well as more insidious disinformation campaigns.

**INTRODUCTION**

It is widely acknowledged by researchers that social bots occupy diverse social media platforms. Defined as software programs that automatically produce content and emulate human behavior, these bots exhibit political and economic influence at an international scale (Ferrara, Varol, Davis, Menczer, & Flammini, 2016). For instance, it has been demonstrated that social bots manipulate political discourse (Bessi & Ferrara, 2016; Conover, Ratkiewicz, & Francisco, 2011; Howard & Kollanyi, 2016; Kollanyi, Howard, & Woolley, 2016; Ratkiewicz et al., 2011), affect stock trading (Fiegerman, 2014; Hwang, Pearce, & Nanis, 2012), and increase the spread of mis and disinformation during emergencies (Gupta, Lamba, & Ponnurangam, 2013).

On February 14, 2018, a mass shooting occurred at Stoneman Douglas High School in Parkland, Florida. The shooter killed 17 people and injured 17 more, making it one of the deadliest mass shootings on record in the US (Follman, Aronsen, & Pan, 2018). Following the shooting, millions of people discussed the event on Twitter. For example, Parkland survivor Emma González, who has been vocally pro-gun control, created a Twitter account following the shooting that has amassed over one million followers as of March 2018.[1]

Within 48 hours of the shooting, social bot monitoring services[2] reported a heightened use of shooting-related hashtags, keywords, and URLs. Specifically, researchers from the group Hamilton68 found that Russian bots were using the #parklandshooting to spread disinformation. The bots advanced unverified claims that the shooter searched for Arabic phrases on Google (Frenkel & Wakabayashi, 2018) and amplified conspiracy theories promoted by far-right outlets that the Parkland survivors were paid crisis actors (Smith, 2018). Presence of bots during the shooting aftermath paralleled documented concerns among journalists (Chen, 2015) and the government (Rid, 2017) of the potential for social bots to cause geopolitical conflict by spreading disinformation and contributing to political polarization.

A week after the shooting on February 21, Twitter released a new set of guidelines prohibiting three types of social bot behaviors enacted via third-party apps: posting similar content to multiple accounts; coordinating simultaneous likes, follows, and/or retweets across various accounts; and automating posts and actions from multiple accounts on a specific topic (Roth, 2018). The company also deleted thousands of accounts that it flagged as social bots. Specific groups, such as right-wing conservatives, criticized Twitter for their resultant loss of followers, their accounts being flagged, and their ability to run ads revoked. Former Twitter employees also criticized the bot crackdown as too little, too late (Lorenz, 2018). They claimed that

---

[1] https://twitter.com/emma4change?lang=en
[2] https://dashboard.securingdemocracy.org/, https://botcheck.me/

Twitter did not prioritize the development of a program sophisticated enough to detect social bots, along with problematic content including spam and abuse (Kosoff, 2018).

We decided to examine the activity of social bots during and following these events for several reasons. Most ostensibly, we considered it a good case because social bots were thought to be active. In addition, a mass shooting provides a new context from which to examine bot strategies, as existing research focuses on elections. Findings from this research, therefore, can provide a contextual counterpoint from which to compare and contrast bot strategies.

Our research meets several objectives. Initial media reporting relied primarily on quantitative evidence (e.g., hashtag frequency of identified bot accounts) to support the claim that bots could incite geopolitical conflict via disinformation campaigns. However, this claim is unsubstantiated without empirical evidence that examines social bot strategies, and which strategies are effective. Therefore, it is possible that we could identify any new changes or developments in bot strategies. Further, we engage in both qualitative and quantitative analysis of social bot strategies and their effectiveness, which provides both micro- and macro-level perspectives on these issues. We applied this analysis to thousands of social bot retweets to address the following research questions: 1) What are the main strategies employed by retweeted social bots? 2) Which strategies are most effective, as indicated by the number of retweets?

**LITERATURE REVIEW**

Multiple, sometimes contesting definitions of social bots exist (Stieglitz, Brachten, Ross, & Jung, 2017). Based on a comprehensive literature review, Stieglitz et al. (2017) define social bots as automated social media accounts that emulate human behavior. In this study, we also adopt this definition. Within the context of Twitter, social bots write original tweets, and follow and retweet other users. Social bots can engage in several malicious activities including identity attacks (e.g., duplicating a celebrity account, impersonating a user's friend or family member to extract personal information from them) (Goga, Venkatadri, & Gummadi, 2015), astroturfing (i.e., making it appear as if the vast majority of people are in favor of a particular position) (Ratkiewicz et al., 2011), smoke screening (i.e., using context-relevant hashtags to distract users from the main points of a debate), and misdirection (i.e., using context-relevant hashtags without referring to the topic at all) (Abokhodair, Yoo, & Mcdonald, 2015). Bots can also be purchased to sway public opinion, such as a politician buying social bots to boost their follower counts (Woolley, 2015). However, social bots can also engage in neutral, or even benign actions, such as chat bots (Stieglitz, Brachten, Ross, et al., 2017). These findings suggest that social bots adopt various strategies depending on the specific objectives of their creators.

Social bots are harder to detect as they become more sophisticated (Everett, Nurse, & Erola, 2016). Further complicating detection are cyborg or hybrid accounts, which are a combination of people and automation techniques (e.g., someone tweeting from their account, but also using a third-party service to tweet similar messages from other accounts) (Chu, Gianvecchio, Wang, & Jajodia, 2012). Researchers have employed three dominant techniques to identify social bots: graph-based, crowdsourcing, and feature-based (Ferrara et al., 2016). Feature-based approaches are the most reasonable for labeling larger datasets as they incorporate network features considered by graph-based models, as well as training data manually labeled by human annotators (Kruschinski, Jürgens, Maurer, et al., 2018; Stieglitz, Brachten, Ross, et al., 2017). Botometer[3] (formerly BotOrNot) represents the only public-facing interface that detects social bots. Several papers have used Botometer to detect bots that are relatively sophisticated (Stieglitz, Brachten, Ross, et al., 2017; Vosoughi, Roy, & Aral, 2018), however bots with higher levels of sophistication (e.g., those that simulate breaks and slightly modify messages) may avoid detection (Stieglitz, Brachten, Berthelé, et al., 2017). Further, Botometer does not detect hybrid or cyborg accounts.

The influence of social bots online is contested within the literature, and varies by context – both of the specific topic investigated (e.g., the US 2016 elections) and how influence is measured. Bessi & Ferrara (2016) found that bots generated approximately 20% of Twitter conversation in the months leading to the 2016 US presidential election. Further, they found that people were equally as likely to retweet other humans as they were social bots. Their findings support those from other work demonstrating the difficulty that Twitter users experience in identifying tweets from social bots, which may be due to the informal and short nature of Twitter messages (Freitas, Benevenuto, Ghosh, & Veloso, 2015). Therefore, there exists a potential for bots to leverage interactions with human users to disseminate misinformation and wage disinformation campaigns via social contagion (Wang, Angarita, & Renna, 2018). Studies of Twitter social bot involvement in other topics, including Brexit (UK) (Howard & Kollanyi, 2016) and political conversations in Venezuela (Forelle, Howard, Monroy-Hernandez, & Savage, 2015) found that social bots played a less prominent role based on content generated and retweeted. However, this role was strategic. For instance, Forelle et al. (2015) found that Venezuela's radical opposition uses the most active bots. Other studies have not found bots to be influential if they are even present in the conversation. For example, Brachten, Stieglitz, Hofeditz, Kloppenborg, & Reimann (2017) found that Twitter conversation around the German 2017 state elections had little social bot activity, the bots demonstrated little to no strategy and little influence as measured by retweets. Recently, Vosoughi et al. (2018) discovered that social bots accelerate the spread of both true and false news, and consequentially, people played more of a role in spreading false news than bots.

---

[3] https://botometer.iuni.iu.edu/#!/

These works illustrate the importance of continued research on social bots as their strategies and influence appear to be constantly changing and highly contextual; indeed, the bots themselves are changing and becoming more difficult to detect. This research contributes to the established body of literature by presenting mass shooting event in the US as a new context.

**METHODS**

*Data Collection*

We chose to collect data following the Parkland shooting for two reasons: 1) prior research demonstrates that political events are rife for the presence of bot accounts, and 2) we wanted to collect recent data from a new context to consider any changes or developments in bot strategies. Following the events, of Parkland, we collected over seven million tweets from over two million unique users using the Twitter Streaming API via Mozdeh (mozdeh.wlv.ac.uk/). We collected tweets from February 16 through March 5, 2018. To collect these tweets, we used nine popular keywords and hashtags related to gun control and the Parkland shooting: *#schoolshooting, #parkland, #guncontrol, #guncontrolnow, #gunreformnow, #nikolascruz, #floridaschoolshooting, #floridashooting,* and *#floridahighschoolshooting*. We defined "popular" based on Twitter trending topics within the first 48 hours of the shooting, as well as hashtags identified by Hamilton68 and Botcheck.me as being wielded by social bots. Once we collected the tweets, we stored them in an SQL database.

*Data Analysis*

From the collected data, we took a random sample of 120,000 Twitter users and ran their usernames through the Botometer classification algorithm to identify bot accounts. The algorithm scored a detection accuracy of 0.95 AUC with ten-fold cross validation when applied to a 2011 dataset of detected social bots (Davis, Varol, Ferrara, et al., 2016). However, this high classification accuracy is likely inflated due to the age of the data. We accessed the Botometer public API endpoint using Python. Botometer's rate limit of 17k accounts per day (40k for a paid account) determined the size of our random sample. Of these 120,000 accounts, Botometer identified 2,739 (2.2%) as having been deleted. It is possible that some of these accounts were bots removed by Twitter. However other instances could represent users deciding to delete their accounts. Botometer classified 5% (6,051) of the remaining 117,261 accounts as social bots.

We then randomly sampled 5,000 social bot accounts from the dataset and queried them in our SQL database to determine how many of these bot accounts were retweeted. We decided to sample from the total number of bots due to the computational expense of a larger query. We found that 11% (557) of the 5,000 social bot accounts were retweeted. Since Botometer could return false positives, particularly flagging organizations like the GOP as bots in our dataset, we then checked our 557 retweeted accounts to determine which were verified by Twitter.[4] We automatically eliminated verified (16.5%, n=92) and deleted accounts (1%, n=6) from our dataset, leaving us with 459 social bot accounts.

In our next step, we queried the SQL database to pull all records that retweeted one of our 459 social bot accounts. We took the Twitter accounts that retweeted our identified social bot accounts (29,269) and ran them through Botometer. Similar to our initial analysis using Botometer, we removed seven deleted accounts and relabeled nine verified accounts initially flagged as bots. In a final step to clean the dataset, we queried our retweet database for any instances where the same Twitter account was flagged by Botometer as both bot and not bot – meaning that Botometer initially flagged it as a retweeted social bot, but the account had also retweeted social bot content captured in our dataset. We found that Botometer inconsistently flagged 51 (11%) of accounts in our dataset. We removed these accounts and their related records from our dataset.

Our final sample consists of 408 social bot accounts retweeted by 19,125 Twitter users, contributing a total of 36,627 retweets. Of these retweets, 1,413 (3.9%) are unique. Botometer flagged the majority of Twitter users retweeting social bot content as non-bots (92.9%, n=17,773) and the rest as bots (7.1%, n=1,352). Therefore, the majority of actors sharing social bot generated content were humans.

We took the sample of 1,413 unique retweets of bot accounts and coded them manually. During this qualitative coding, we sought to identify social bot strategies, which we defined as the inferred goals or aims of the social bot activity based on the type of content shared. We constantly compared emerging codes to existing ones, revising and combining coding categories (Charmaz, 2014). We ultimately created a codebook that nested types of content shared underneath higher-level bot strategies. Once we established a preliminary codebook, two members of the research team coded 20% of the data and discussed coding discrepancies, revising the codebook as necessary. We calculated reliability among two coders using percentage agreement at 88%. Following coding, we wanted to determine which strategies were most effective, which we operationalized by the number of retweets. To determine this number, we took our unique tweets and ran an SQL query to count the number of times they were retweeted in our dataset. Our dataset can be accessed at the following link: https://doi.org/10.6084/m9.figshare.6726419.

---

[4] https://help.twitter.com/en/managing-your-account/about-twitter-verified-accounts

**FINDINGS**

We identified six key strategies employed by retweeted bot accounts. These strategies are: Baiting, Instilling Public Doubt, Sharing Information, Spreading Conspiracy Theories, Organizing Political Action, and Using the Shooting for Commercial Gain. We now discuss each strategy in more detail. From this discussion, we exclude the 8% (n=2,944) we coded as irrelevant (e.g., tweets about a basketball game in Florida). Below, we only use examples of tweets still publicly available on Twitter. However, it should be noted that since Twitter has recently started to shut down accounts suspected to be social bots, it is possible that these tweets will not be available by the time of publication of this paper. We have removed the names of all Twitter accounts, including bot accounts (since Botometer is not likely 100% accurate), to protect user privacy. Table 1 displays a list of all high-level strategies, their definitions, and frequencies.

| Main Code | Definition | N | % (of 36,628) |
|---|---|---|---|
| Baiting | Instances where the tweet content elicits an angry or emotional response. | 21,453 | 58.6% |
| Instilling doubt | Posting dubious information that makes people doubt democratic institutions | 6,566 | 17.9% |
| Sharing information | Sharing factual information about the event | 4,046 | 11.0% |
| Spreading conspiracy theories | Suggesting that the shooting event was part of a secret plan to do something unlawful or harmful | 1,458 | 4.0% |
| Organizing political action | Attempting to elicit participation in a political event | 88 | 0.2% |
| Using shooting for commercial gain | Attempting to profit from the shooting | 73 | 0.2% |

Table 1. High-level Social Bot Strategies

*Baiting*

The majority of retweets showcased the social bot strategy of baiting. A popular example of baiting was social bots retweeting content that criticized key actors in the Parkland shooting (36.6%, n=13,424). Consider the following tweet, which criticizes the media outlet CNN:

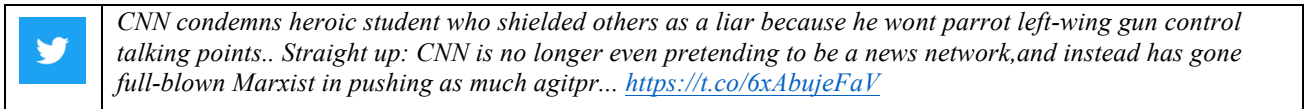

*CNN condemns heroic student who shielded others as a liar because he wont parrot left-wing gun control talking points.. Straight up: CNN is no longer even pretending to be a news network,and instead has gone full-blown Marxist in pushing as much agitpr... https://t.co/6xAbujeFaV*

This bot-generated tweet shares the first 280 characters of an article from a blog, Natural News (www.naturalnews.com) that sells various dietary supplements and alternative medicine, as well as shares false scientific news and conspiracy theories. Such content is standard for bot accounts, which predominately share content generated by others. This content represents a baiting strategy because of the language used. Connecting CNN to Marxist ideologies can elicit an emotional response among those who already mistrust the news network and believe that its coverage disproportionately represents a specific political agenda. Other actors frequently criticized in bot retweets were conservative politicians (10.1%, n=3,682), the FL police officers responding to the shooting (5.9%, n= 2,169), and liberal politicians (5.1%, n=1,862).

Another popular baiting strategy was bots' sharing of content that either was explicitly against gun control (4.6%, n= 1,672) or for gun control (3.4%, n=1,255). Similar to the prior example, this sort of content represented retweets or excerpts of news content that had a distinct opinion on the gun control issue. The following tweet exemplifies a pro-con control strategy:

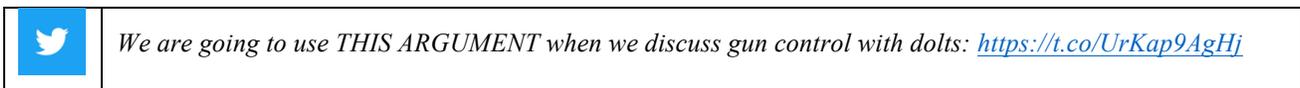

*We are going to use THIS ARGUMENT when we discuss gun control with dolts: https://t.co/UrKap9AgHj*

Here, the use of the word "dolts" to describe individuals against gun control represents baiting. It can both incite those who agree with gun control measures to share this content, as well as potential replies from those who disagree with gun control and are insulted by being called a "dolt."

Two other baiting strategies of note are drawing false equivalencies (3.9%, n=1,444) and inviting debate on gun control (2.2%, 794). Drawing false equivalencies represents when bots share content that equates mass shootings to some other controversial issue. For instance, the following example positions Planned Parenthood as more dangerous than school shootings:

> 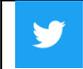 *Planned Parenthood has killed more babies last year than all school shooting COMBINED! #stopplannedparenthood*

Drawing false equivalencies parallels smoke-screening strategies identified in prior work (Abokhodair et al., 2015), because bots use content with relevant hashtags and keywords representative of a particular issue (here, gun control) to ultimately distract individuals by having them discuss another topic. In the majority of cases, the issue meant to distract from a broader discussion surrounding mass shootings is societally contentious (e.g., abortion).

Inviting debate on gun control is a strategy used by bots to bring individuals from multiple sides of an issue into one Twitter thread, as illustrated by the following example. In this example, the Twitter account uses hashtags visible to individuals who stand on multiple sides of the gun control debate, asking them to retweet to "vote" on the issue. Ultimately, this strategy spreads this tweet to more Twitter accounts and can also artificially lend support to a specific stance on the issue based on retweet volume.

> 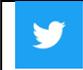 *Should teachers be armed in schools? Retweet for more votes! #NRA #GunControlNow #GunReformNow*

### *Instilling Doubt*

Instilling doubt was the second-most popular category after baiting. Content within this category focused on the media (9.4%, n=3,456), FBI (4.9%, n=1,806), and the police (2.8%, n=1,034). The following tweet illustrates instilling doubt in the media:

> 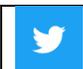 *FL School Shooting Survivor Says CNN Gave Him a Scripted Question For Townhall Meeting https://t.co/BwOdiuXonr*

Here, the bot retweets a headline and link to a story from enVolve (https://en-volve.com/), a conservative news and opinion website that has been identified by fact-checking organizations as a source that creates and shares false news (Politifact, 2017). Several fact-checking organizations have debunked the story about CNN handing a scripted question to a Parkland survivor ("Did CNN give a Parkland shooting survivor scripted questions?," 2018). Therefore, this story goes one step beyond critique by instilling doubt about CNN's reporting practices by making a false claim about the network.

Instilling doubt in the FBI and police are connected. The majority of tweets coded under this category shared dubious information to insinuate that both institutions aided the shooting. For instance, the following tweet blames the FBI for ignoring tips about the shooter to focus on the Russia investigation:

> 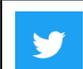 *Anyone else ticked about the FBI wasting resources on a phony witch hunt while letting 17 people get slaughtered. They were WARNED They knew his complete NAME: Nikolas Cruz They LET this happen on purpose But we're wasting MILs to appease Libs who're mad they lost an election?*

Similar to baiting, this tweet also elicits anger, both against the FBI and liberal politicians, who are portrayed by the tweet as letting a shooting happen for political gain. The dubious information in this tweet is twofold – that the Russia investigation is a "phony witch hunt" and that the FBI purposefully let the shooting happen. Tweets instilling doubt in the police follow a similar pattern but insinuate that the police worked to abet the shooting. For instance, one bot tweeted the following:

> 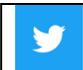 *BREAKING! EMS Claims Broward Sheriffs Office Blocked Them From Entering FL School During Shooting https://t.co/znIlkeeABy*

The article to which the bot links is another false news site, Clash Daily (clashdaily.com) (Politifact, 2017). This headline likely derives from the "stand down" conspiracy theory, which purported that the police gave EMS a "stand down" order during the shooting. What prevented us from coding this tweet as a conspiracy theory is that it does not cite a "stand down" order, however it does misrepresent standard EMS protocol to wait for police affirmation before entering the scene of a shooting ("Did a first responder say he was told to 'stand down' in Parkland?," 2018). Therefore, the tweet makes a dubious claim about the Broward County Sherriff's office that could make a reader question the intents and motivations of police departments during mass shootings.

### *Sharing Information*

A small, but significant proportion of bot retweets shared information. The following tweet illustrates one example of information sharing:

> 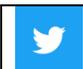 *Russian Bots Moved Quickly to Exploit the Florida Shooting - The New York Times #DemForce https://t.co/ippBU4Gr0g*

Presence of these tweets confirms findings from other research that not all social bot accounts are malicious. Benign bots also exist, particularly in journalistic contexts to share news stories (Lokot & Diakopoulos, 2016). Interestingly, bot accounts that engaged in the spread of more malicious information (e.g., instilling doubt) could also spread information. Consider the following example from a bot account that predominantly tweeted malicious tweet content:

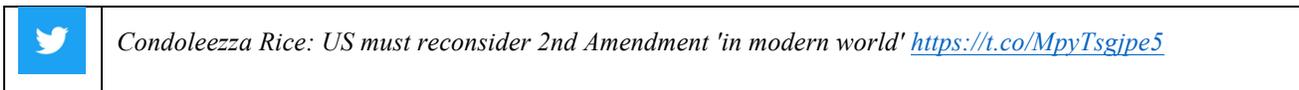
*Condoleezza Rice: US must reconsider 2nd Amendment 'in modern world' https://t.co/MpyTsgjpe5*

This account is of particular interest, as it links to a site called http://ytreet.com (see Figure 1 below). A small, but significant number of bot tweets (1.2%, n=439) originated from this domain http://ytreet.com/ye. This latter link directs to a log-in page (see Figure 2). This log-in page suggests a potential botnet or coordinated program that communicates across multiple devices to complete a task. Therefore, this finding suggests that a single bot account can engage in various strategies, which can be aided by botnets.

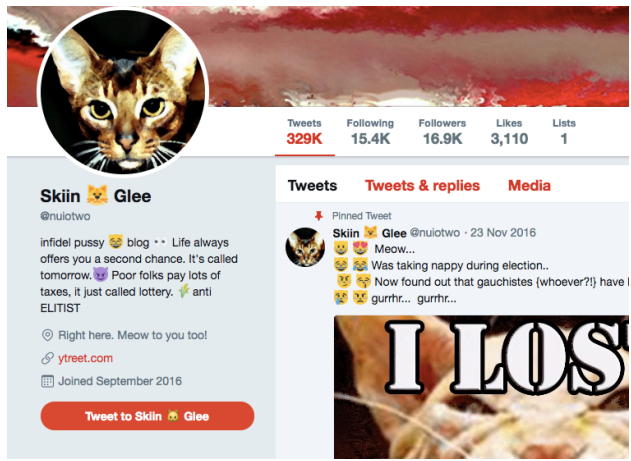
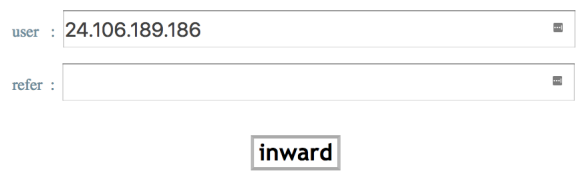

**Figure 1.** Suspected social bot account. Note that it would be impossible for a human to have tweeted 329,000 tweets in 18 months. The account tweets an average of over 500 tweets a day.

**Figure 2.** Log-in page to the suspected botnet.

### *Spreading Conspiracy Theories*

This bot strategy confirms media reports of suspected bot activity and motivations following the Parkland shooting. The most popular conspiracy theory included in retweeted bot content was the "false flag" theory (Starbird, Maddock, Orand, Achterman, & Mason, 2014). This theory purports that the government planned the shooting and aided by the police and media to justify gun control measures. Below is an example of a tweet supporting a false flag conspiracy theory:

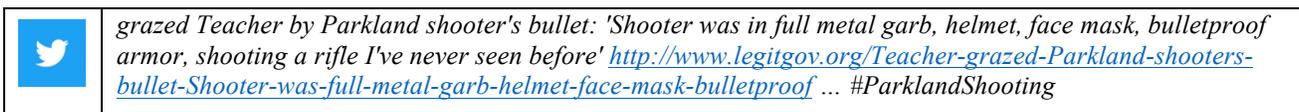
*grazed Teacher by Parkland shooter's bullet: 'Shooter was in full metal garb, helmet, face mask, bulletproof armor, shooting a rifle I've never seen before' http://www.legitgov.org/Teacher-grazed-Parkland-shooters-bullet-Shooter-was-full-metal-garb-helmet-face-mask-bulletproof ... #ParklandShooting*

Here, the account links to a news blog, LegitGov (www.legitgov.org) that shares an alternate narrative of a shooting event. This story insinuates that the shooting was planned by questioning the accepted narrative of who the shooter was, how he was dressed, and what equipment he had. Another example of a popular conspiracy theory relates to the accusing Parkland shooting survivors as being crisis actors. For instance, the following tweet accuses survivor, David Hogg, of being coached by the FBI:

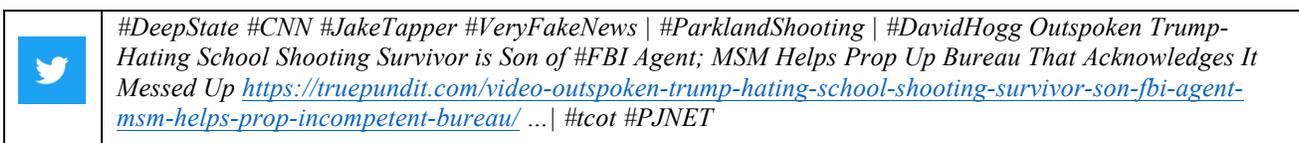
*#DeepState #CNN #JakeTapper #VeryFakeNews | #ParklandShooting | #DavidHogg Outspoken Trump-Hating School Shooting Survivor is Son of #FBI Agent; MSM Helps Prop Up Bureau That Acknowledges It Messed Up https://truepundit.com/video-outspoken-trump-hating-school-shooting-survivor-son-fbi-agent-msm-helps-prop-incompetent-bureau/ ...| #tcot #PJNET*

The post links to True Pundit (truepundit.com), another blog sharing conspiracy theories and false news (Politifact, 2017). What many of these conspiracy theories have in common is that they take advantage of the confusion inherent to crisis events that may influence the presence of contradictory accounts, such as a witness claiming the shooter was heavily armed and outfitted. Further, they take advantage of facts, like that David Hogg's father is a retired FBI agent, to "prove" unsubstantiated claims about the shooting being a government-led conspiracy.

*Political Organizing and Using Shooting for Commercial Gain*

A smaller subset of tweets engaged in political organizing and using the shooting for commercial gain. Although this category is important because it could lead to political participation, as a bot strategy it was the least frequently retweeted. This lack of retweets does not reflect the reality that people engage in political organizing online, only that social bots may not have played a significant role in facilitating such organization within the context of the Parkland shooting. Another reason for the lack of retweets may have been the tweet collection time period, which ended as the March for our Lives walkouts were beginning. In the context of tweets analyzed, political organizing mostly encouraged participation, such as going to a walkout or boycotting an organization. The following tweet links to an article encouraging individuals to boycott companies sponsored by the NRA:

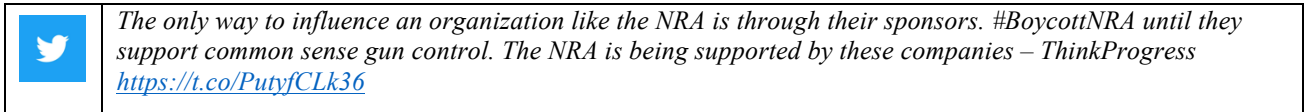

*The only way to influence an organization like the NRA is through their sponsors. #BoycottNRA until they support common sense gun control. The NRA is being supported by these companies – ThinkProgress https://t.co/PutyfCLk36*

Using the shooting for commercial gain was used by some to make money from the shooting. For instance, the following account tweeted links to buy anti-gun protest t-shirts:

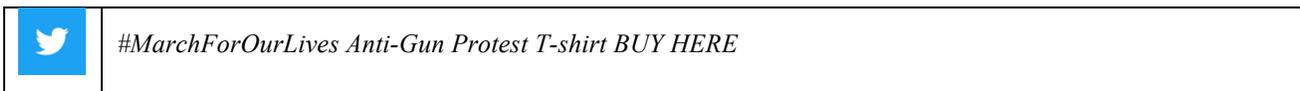

*#MarchForOurLives Anti-Gun Protest T-shirt BUY HERE*

Again, the number of retweets for this bot strategy may have been low due to the data collection time period or simply that this bot strategy, while it exists, is not particularly effective as evidenced by number of retweets.

**DISCUSSION**

Our research examined retweeted content generated by social bots following a mass shooting event. We used a mixed method approach to collect and analyze our data, facilitating both macro- and micro- perspectives of bot strategies during this event. Findings indicated that the majority of actors retweeting bot accounts were humans. This finding supports the conclusions of prior work that while bots *amplify* conversation surrounding significant issues (e.g., political elections, crisis situations), people are primarily responsible for the dissemination of bot-generated content (Bessi & Ferrara, 2016). Another significant finding was that a subset of bot-generated content is not malicious. This content could be produced by "benign" bots, employed within journalism contexts to share news related to the Parkland shooting and its aftermath (Lokot & Diakopoulos, 2016). Therefore, reporting by media outlets on the presence of bot activity following the Parkland shooting may be substantiated by this analysis, but with these two critical caveats.

We coded the most popular social bot strategy as baiting. The cultural theory of risk can potentially explain the popularity of baiting content in our dataset (Douglas & Wildavsky, 1982). This theory contends that people have sophisticated social and cognitive reasons for seeking, sharing, and using specific information, even if that information is misleading or false. People develop risk perceptions based on their commitment to a particular way of life and worldview. Therefore, people may share content that criticizes institutions like the mainstream media if they view this institution as presenting a risk to their way of life. Content that bolsters a specific ideology, such as a pro-gun control agenda, may appeal because it affirms their sense of self. This sense is derived, in part, by the groups to which people belong. Therefore, supporting specific in-group ideologies and members by sharing content may help reinforce an individual's perception that they belong to the "correct" group with desirable ideologies (Sherman, Cohen, & Sherman, 2002).

Instilling doubt and conspiracy theories were two other key forms of bot-generated content that we found during data analysis. These two categories share patently false information and reflect popularized fear of bot activities discussed in the media. Prior research has established that the primary motivations for circulating this content, at least among bots, is to undermine trust – both in information and democratic institutions (Pomerantsev & Weiss, 2014; Starbird, 2017). We found that some bot accounts appeared to share a variety of content – ranging from conspiracy theories to more benign information. Further, there are likely botnets at work in disseminating this content, meaning that it is coming from a variety of sources, which can inaccurately foster the perception among people that these fringe ideas are more mainstream than in actuality.

Similar to the discussion on baiting, people are not necessarily cognitively or culturally helpless when processing and evaluating questionable or false information. Instead, these individuals may have a "crippled epistemology" from being exposed to limited media sources (Sunstein & Vermeule, 2008; see also Starbird, 2017). We found evidence for this perspective in qualitative analysis, where much of the content linked to sources that parroted similar narratives, most often supporting a false flag perspective.

These observations suggest the need to reframe our approach to issues like information literacy. As boyd (2018) argues, people often employ critical assessment and analysis of information sources. However, these standards of assessment and analysis vary based on cognitive and cultural perspectives and worldviews. For example, someone that retweets content instilling doubt in the media recognizes media strategies such as framing (e.g., questioning CNN's support of pro-gun control narratives), even if they are couched within a more substantial, false narrative (e.g., CNN supported its pro-gun agenda by feeding scripted questions to Parkland survivors). Therefore, addressing the spread of false information cannot be solely solved by teaching information literacy and critical thinking. Instead, it is important to recognize people's motivations for seeking, sharing, and using particular information and how the strategies used by actors, including bots, appeal to these motivations. Some concrete recommendations for improving information literacy instruction based on these findings include:

- **Outline approaches for identifying social bots.** Common heuristics, such as a high number of retweets for a recent account, indicate a high likelihood of a bot account (Ferrara et al., 2016). Individuals can also insert a Twitter account name into the Botometer interface to determine the likelihood of it being a social bot.
- **Overview common strategies used by social bots.** This paper offers a beginning typology, which can be combined with strategies identified by prior research, such as smoke screening and misdirection (Abokhodair, Yoo, & Mcdonald, 2015). Have individuals question whether there are underlying strategies present in social media content they encounter and what reactions these strategies are trying to elicit from them.
- **Engage in a dialog about media framing**. Have individuals share common frames they experience when engaging with media. What are some frames that individuals may have blind spots towards based on their underlying ideological views?
- **Do not discount emotions and group think.** Have an open discussion of how emotions and group memberships shape how individuals interpret and share information on social media. Stress the importance of striving for accuracy as defined by an information literacy framework and attempt to re-motivate individuals to strive for accuracy goals.

**CONCLUSION**

In this study, we examined the strategies of social bots whose content was shared on Twitter on the topic of gun control following the school shooting in Parkland, FL. Analysis of over 36k tweets from more than 400 social bot accounts identified six key bot strategies. Findings indicate that other people, rather than bots primarily shared bot-generated content. Reasons for the effectiveness of these strategies, as conveyed by the number of retweets, can be attributed to cultural and cognitive motivations that people have to protect their worldviews and related conceptions of self-worth.

Limitations of this work relate to data collection and analysis methods. For data collection, we used the Twitter stream, which collects a random sample of 1% of all Twitter data shared within a 24-hour period. Therefore, our data is not necessarily representative of all Twitter activity centered on the Parkland shooting. Further, we started the Twitter stream two days after the Parkland shooting and are therefore missing tweets from the outset of the event. This lack of sampling is somewhat countered since some content shared during these first two days was widely retweeted for days following the event, meaning that it is included in our dataset. However, to get a better picture of characteristics such as what bot content was shared and when our future work will incorporate this historical data.

Our data analysis was limited by the Botometer algorithm in detecting social bot accounts. We noted that Botometer inconsistently flagged 11% of Twitter accounts as bots when we ran these accounts through the algorithm twice, a few days apart. Further, Botometer flags institutional accounts as social bots, such as the news site, Vice. While we eliminated the inconsistently flagged and verified institutional accounts from the data, it is likely that some of the accounts identified as bots in the data are not bots. This limitation reflects the increasing sophistication of social bot accounts and the need for media companies like Twitter to continually provide researchers with training data to flag these accounts accurately over time. In addition, Botometer was trained on 2011 data, rendering its accuracy (0.95 AUC) as likely lowered due to this increasing sophistication of bot accounts. However, as of March 20, 2018 (after we conducted this research), the Botometer classifier was updated with thousands of new data points, which may have improved the accuracy of the classifier.[5] We will use this new version of Botometer in future work.

Since our research is exploratory, there are many directions for future work. Our next step with this data is to fold in historical tweets from the first days of the shooting and incorporate retweets of all flagged bot accounts. Due to Botometer's limitations, we plan to implement additional methods for verifying social bot accounts, including further feature-based identification of false positives and negatives, as well as using manual coding. The initial coding scheme for bot strategies can be expanded and refined to include other contextual elements, such as political ideologies (e.g., liberal, conservative) of those retweeting the content. We can use the labeled data to train a classifier that can code a larger sample of the data to determine how these strategies are disseminated across all collected tweets. Additionally, we can incorporate additional methods of analysis, including network analysis, to further describe the dissemination of tweet content. It would also be of interest to look at other Twitter activity, such as liking or replying, to see how bot strategies align with or differ from retweets. Finally, we will also sample from other events and platforms to track identified bot accounts and determine how their strategies may vary by context.

---

[5] https://botometer.iuni.iu.edu/#!/faq

In sum, our research contributes to a growing body of work describing the strategies of social bots in online environments like Twitter. Understanding these strategies can inform efforts by other researchers and policy organizations related to the combat of misinformation, as well as more insidious disinformation campaigns.